\documentclass{article}

\usepackage{microtype}
\usepackage{graphicx}
\usepackage{subcaption}
\usepackage{booktabs} 

\usepackage{hyperref}

\usepackage[accepted]{icml2026}

\usepackage{amsmath}
\usepackage{amssymb}
\usepackage{mathtools}
\usepackage{amsthm}
\usepackage{listings}
\usepackage{xcolor}
\lstdefinelanguage{pvs}{
    basicstyle=\ttfamily\small,
    keywordstyle=\color{blue}\bfseries,
    commentstyle=\color{gray},
    stringstyle=\color{teal},
    showstringspaces=false,
    breaklines=true,
    frame=single,
    backgroundcolor=\color{gray!10},
    morekeywords={
        THEORY, BEGIN, END, IMPORTING, LIBRARY,
        bool, TRUE, FALSE
    },
    sensitive=true
}

\lstdefinelanguage{json}{
    basicstyle=\ttfamily\small,
    numbers=left,
    numberstyle=\tiny,
    stepnumber=1,
    numbersep=8pt,
    showstringspaces=false,
    breaklines=true,
    frame=single,
    backgroundcolor=\color{gray!10},
    stringstyle=\color{blue},
    commentstyle=\color{gray},
    keywordstyle=\color{red},
    morestring=[b]",
    morecomment=[l]{//},
}

\usepackage[capitalize,noabbrev]{cleveref}

\theoremstyle{plain}

\theoremstyle{definition}

\theoremstyle{remark}

\usepackage[textsize=tiny]{todonotes}

\icmltitlerunning{Submission and Formatting Instructions for SD4H Workshop at ICML 2026}

\begin{document}

\twocolumn[
  \icmltitle{Formally Verified Code Synthesis for Structured Data Translation in a \\ Medical Internet of Things}



  \icmlsetsymbol{equal}{*}

  \begin{icmlauthorlist}
    \icmlauthor{Colin Samplawski}{yyy}
    \icmlauthor{Adam D. Cobb}{yyy}
  \end{icmlauthorlist}

  \icmlaffiliation{yyy}{Computer Science Laboratory, SRI International}

  \icmlcorrespondingauthor{Colin Samplawski}{colinsamplawski@gmail.com}

  \icmlkeywords{Machine Learning, ICML}

  \vskip 0.3in
]



\printAffiliationsAndNotice{}  

\begin{abstract}
  In this work we present a LLM powered, evolutionary code synthesis system for structured data translation in a Medical Internet of Things settings. A key challenge in this domain is ensuring that the synthesized code is trustworthy and reliable. To this end, we integrate a formal verification step into our code synthesis pipeline to ensure that any generated code is guaranteed to satisfy predefined requirements. In particular, we present a case study of integrating a novel device (a pulse oximeter) into the existing network of devices. Our system generates a formally verified translation between the device's JSON schema and the  Fast Healthcare Interoperability Resources (FHIR) format used by the wider system. This formal verification stage ensures structured data translated by the generated code will always be in the target output schema. We provide a set of experimental results which demonstrate that our system is able to consistently generate correct translation at low cost. 
\end{abstract}
\section{Introduction}
Large language model (LLM) powered code synthesis has become an increasingly important tool for modern software development \cite{jalil2025transformative}. While these tools can enable significant gains in productivity, they also introduce significant risk and novel challenges. LLMs are well known to hallucinate and can introduce subtle vulnerabilities into the code \cite{kalai2025language}. These problems are further compounded by the shear scale and speed of LLM code generation, making manual human verification of critical code laborious and intractable. These risks are especially acute in safety-critical domains, such as a healthcare.


In this work, we introduce a code synthesis pipeline that formally verifies synthesized code for a  Medical Internet of Things (MIoT) network \cite{dimitrov2016medical}. This work is part of a broader project with the goal of reconfiguring a vehicle with medical devices, such that the vehicle is primed to deliver mobile care. A key challenge of this project is the integration of novel medical devices to the vehicle's MIoT network over time, without assuming any available (human) software development expertise. Furthermore, given the safety-critical nature of the system, any automatically synthesized code must be performed in a manner that ensures reliability and trust. 

To tackle this problem, we follow a neuro-symbolic architecture. We use a neural module that uses LLMs as operators within an evolutionary algorithm \cite{lange2025shinka} and a theorem prover module which we have adapted to JSON schema verification \cite{owre1999formal}. This architecture ensures that the significant reasoning power of the LLM is grounded by the formal theorem prover. 
As a proof of concept, we present a case study of using our proposed system to integrate a pulse oximeter into our MIoT environment. We translate the device's provided JSON schema to the more standard Fast Healthcare Interoperability Resources (FHIR) \cite{vorisek2022fast}. We perform a set of experiments to analyze the success rate and resource cost of using our approach to generate a formally verified translation in this setting. We find that our system is able to consistently generate correct translations at low cost.

This rest of this paper is structured as follows. In Section \ref{sec:background} we discuss background work and motivation. Then we introduce our approach and summarize its constituent parts in Section \ref{sec:approach}. In Section \ref{sec:results} we present results of our pulse oximeter case study. Finally we provide an overview of future directions and conclude in Section \ref{sec:conc}.

\section{Background} \label{sec:background}
\subsection{Evolutionary Algorithms with LLMs}
Significant recent work has considered using LLMs as the primary operators inside evolutionary algorithms for code optimization and scientific discovery \cite{novikov2025alphaevolve,openevolve,lange2025shinka}. In these systems, the LLM replaces traditional mutation and crossover operators by proposing semantically meaningful program edits, design variations, or hypotheses conditioned on prior search history. This approach is appealing because LLMs can exploit structure in code and scientific text, enabling searches over large, discrete spaces that would be difficult to explore with purely random or hand-designed operators. At the same time, the flexibility of LLM-generated candidates makes reliability, constraint satisfaction, and evaluation efficiency central challenges for practical deployment.

\subsection{Formal Verification of Synthesized Code}
As generative models are increasingly used to synthesize executable programs, ensuring correctness has become a central concern. Formal verification provides a principled framework for checking whether generated code satisfies specifications such as type safety, functional correctness, invariants, or resource bounds, rather than relying only on example-based testing \cite{herklotz2021formal}. In synthesis pipelines, verification can be used either as a post hoc filter to reject invalid candidates or as part of the search objective itself, guiding generation toward programs that are provably correct. This is especially important in high-stakes domains, where small implementation errors can lead to unsafe or clinically unacceptable behavior.

\subsection{Healthcare Data Interoperability}
Healthcare data interoperability concerns the ability of heterogeneous systems to exchange, interpret, and use structured clinical information consistently across institutional and software boundaries \cite{iroju2013interoperability}. Modern interoperability efforts are shaped by standards such as HL7 FHIR \cite{vorisek2022fast}, standardized clinical terminologies, and structured exchange formats that aim to make patient records, observations, medications, and care plans computable across platforms. Despite this progress, real-world healthcare data remain fragmented, inconsistently coded, and deeply embedded in local workflows, creating substantial barriers to reliable automation and downstream analytics \cite{mathew2015big}. As a result, methods that generate or transform healthcare software artifacts must account for both syntactic conformance to standards and semantic alignment with clinical intent.

\begin{figure}
    \centering
    \includegraphics[width=0.86\linewidth]{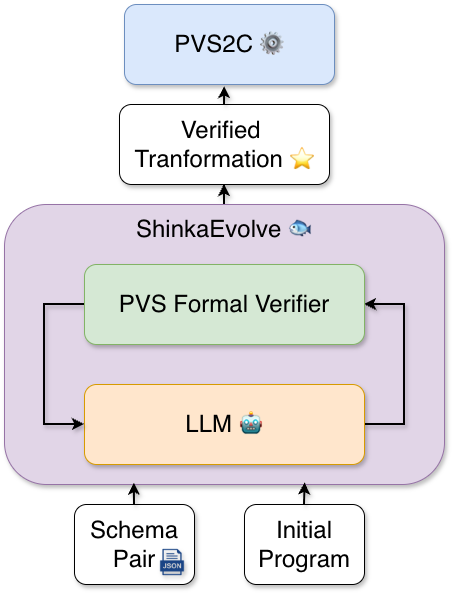}
    \caption{Overview of our proposed system. A user provides an input/output schema pair and initial PVS program. The ShinkaEvolve loop elicits LLM translation code and formally verifies them. Upon successful verification the resulting theorem can be then be translated into C code for use in the MIoT environment.}
    \label{fig:approach}
\end{figure}

\section{Approach}\label{sec:approach}
Our approach, shown in Figure \ref{fig:approach}, is built on ShinkaEvolve \cite{lange2025shinka}. ShinkaEvolve is an open-source implementation of recent evolutionary code optimization techniques leveraging LLMs. ShinkaEvolve provides a multi-island evolutionary approach which builds an SQL database of programs and corresponding fitness scores. At each stage, a parent program is selected along with an LLM sampled from a model ensemble. Additionally, a set of $K$ inspiration programs are selected from the same island. The parent and inspiration programs, along with their fitness scores, are then provided to the LLM as a prompt. The LLM is then instructed to choose between editing, re-writing, or mutating earlier programs to generate a new solution candidate. This candidate is then scored by the provided evaluator and its results recorded into the database. This process then continues in a multi-worker asynchronous loop, working to maximize the fitness function.

In our approach, we use the Prototype Verification System (PVS) as our evaluator for determining solution fitness. PVS is a theorem prover designed to support the formal specification and verification of complex systems \cite{owre1999formal}. PVS provides a domain-specific language (DSL) which we adapt to support the definition of transformations between valid JSON schemas. The pipeline then requires an input schema and an output schema. The input schema determines the format of the incoming data stream (e.g. from some medical device) and the output schema determines the required format of the outgoing data stream (e.g. the patient database in the larger system).

A theorem is then over a translation between inputs of the input schema type and outputs of the output schema type. The transform is correct only if this theory is proven. This is formalized as a set of constraints and core verification condition, which is shown in Listing \ref{lst:pvs-constraints}. Here we define \texttt{input\_compliant?}, \texttt{output\_compliant?}, and \texttt{transform} to be the JSON input schema, JSON output schema, and the synthesized theorem respectively.

In the evolution loop, the LLM is tasked with writing such a transform in the PVS DSL. In the evaluation phase, we then can formally verify if the generated theorem is true. This results in a binary fitness function: 0 for an invalid theorem and 1 for a valid theorem. We note that a challenge of our approach is that the the DSL of PVS is only weakly known by the pretrained LLMs (i.e. the GPT family of models \cite{singh2025openai}). For this reason we additionally provide a PVS language reference and an in-context example transform demonstrating translations between a pair of simple schemas in the prompt.


After this evolution process is complete, we will have a formally verified translation from one JSON schema to another. However, the DSL of PVS does not provide code that can be operationalized. We then use additional tooling which allows us to deterministically transform the verified theorem into C code which can efficiently perform the translation execution inside the MIoT.

\begin{lstlisting}[language=pvs, caption={Constraints and core verification condition}, label={lst:pvs-constraints}]
A_B_constraints: THEORY
BEGIN
  jsltlib: LIBRARY = "../../jslt_pvs/jsltlib/"
  IMPORTING jsltlib@jsondata

  requires(obj: jsondata): bool = TRUE
  ensures(obj, new_obj: jsondata): bool = jdict?(new_obj)
END A_B_constraints

...

A_B_verif: THEORY
BEGIN
  IMPORTING A_B_constraints
  % plus imports/definitions for input_compliant?, output_compliant?, and transform

  correctness: THEOREM
    FORALL (obj: input_compliant):
      requires(obj) IMPLIES
        LET new_obj = transform(obj) IN
          output_compliant?(new_obj) AND ensures(obj, new_obj)
END A_B_verif
\end{lstlisting}





\section{Case Study: Integration of a Pulse Oximeter}\label{sec:results}
In this section we provide an analysis of a concrete use case of our system for integrating a new device (a pulse oximeter) into the already existing MIoT environment. The device comes with a custom JSON schema which encodes a measurement of a patient's pulse and oxygen saturation (shown in Appendix Listing \ref{lst:input_schema}). To be integrated into the larger system, this data must be converted into the Fast Healthcare Interoperability Resources (FHIR) format (shown in Appendix Listing \ref{lst:output_schema}), which is a standardized medical format schema that the system already natively understands.

As input to the ShinkaEvolve process we provide these two known schemas as well as a initial skeleton program which is shown in Listing \ref{lst:pvs-init}. This initial program simply returns an empty JSON dictionary for any input. To rigorously analyze the overall performance of ShinkaEvolve for this task, we perform 10 separate experimental runs for this task. We use two separate islands and run each trial for a total of 20 generations. A successfully verified PVS translation is shown in Appendix Listing \ref{lst:correct_pvs}.





\begin{lstlisting}[language=pvs, caption={PVS theory initial program}, label={lst:pvs-init}]
A_B: THEORY
BEGIN
  jsltlib: LIBRARY = "../../jslt_pvs/jsltlib/"
    
  IMPORTING jsltlib@jsondata

  transform(i: jsondata): jsondata = jdict(())

END A_B
\end{lstlisting}

\subsection{Aggregate Performance}
Table~\ref{tab:aggregate} summarizes key performance metrics averaged across all 10 experimental runs. The earliest successful theorem occurs at generation $5.0 \pm 4.3$ on average with a 
range of 1--14. We define the success rate to be fraction of LLM-generated programs that pass verification across any generation. For this example we obverse a success rate of $37.4 \pm 22.4\%$, reflecting the difficulty of synthesizing correct PVS transform functions.

The average total API cost per experiment is $\$0.87 \pm \$0.18$, but the cost incurred before the first successful proof is only
$\$0.18 \pm \$0.19$, roughly 20\% of the total budget. The remaining cost is then spent on continued exploration after a solution has already been found. The total wall-clock time per experiment averages $42.7 \pm 8.1$~minutes, with the the LLM generation time taking up the majority of this time. We find that the PVS verification step is generally fast compared to the LLM generation phase. We observe that the first successful program is found in just $10.4 \pm 9.9$~minutes on average.

\begin{table}[t]
  \centering
  \caption{Aggregate experiment statistics across 10 independent runs (20 generations each).}
  \label{tab:aggregate}
  \small
  \begin{tabular}{lr}
    \toprule
    Metric & Mean $\pm$ Std \\
    \midrule
    Earliest Successful Generation & $5.0 \pm 4.3$ \\
    Success Rate & $37.4 \pm 22.4\%$ \\
    Total Cost & $\$0.87 \pm 0.18$ \\
    Cost to First Success & $\$0.18 \pm 0.19$ \\
    Total Wall-Clock Time (mins) & $42.7 \pm 8.1$ \\
    Wall-Clock Time to First Success (mins) & $10.4 \pm 9.9$ \\
    \bottomrule
  \end{tabular}
\end{table}

\begin{figure}
    \centering
    \includegraphics[width=1.0\linewidth]{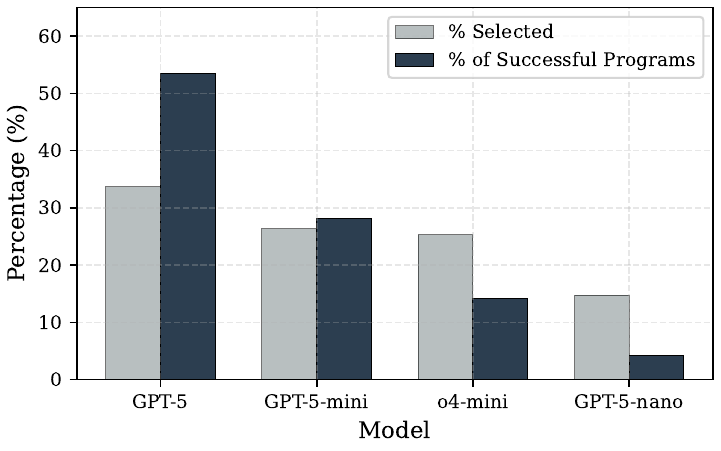}
    \caption{Selection and success rate of LLMs in the ensemble.}
    \label{fig:llm_selection}
\end{figure}
\subsection{LLM Selection}
In our experiments, ShinkaEvolve is configured to use an ensemble of GPT LLMs: \textsc{o4-mini}, \textsc{gpt-5}, \textsc{gpt-5-mini}, \textsc{gpt-5-nano}. At each generation, a member from the ensemble is randomly selected to generate a new solution candidate. The weights of this sampling process are set dynamically using a Thompson sampling bandit approach \cite{daniel2018tutorial} based on observed success from prior generations. Figure~\ref{fig:llm_selection} compares each model's selection frequency against its share of successful programs across all experiments. \texttt{gpt-5} accounts for 33.7\% of all
model selections but is responsible for 53.5\% of successful
programs. Conversely, \texttt{o4-mini}
(25.3\% selected, 14.1\% of successes) and \texttt{gpt-5-nano} (14.7\%
selected, 4.2\% of successes) underperform relative to their selection
frequency. \texttt{gpt-5-mini} performs roughly in proportion (26.3\%
selected, 28.2\% of successes).

This demonstrates that the larger and more expensive LLMs, such as \texttt{gpt-5}, are in general more accurate and therefore favored in the sampling process. We note that this sampling algorithm in ShinkaEvolve considers only an LLMs success rate and not its resource cost.

\subsection{Solution Convergence}
A notable finding is that all 10 experiments converge to essentially the same solution despite following different LLM selection paths and encountering different intermediate failures. Every best-of-run program produces the same FHIR Observation resource structure. Syntactic variations exist across runs, but the semantic content is identical. This convergence suggests that the PVS verification constraint is highly prescriptive: the output schema, combined with the proof obligations, admits essentially one valid mapping from the input. The evolutionary process thus functions less as an open-ended search and more as an oracle-guided synthesis procedure. 

\section{Conclusion}\label{sec:conc}
We have built a fully integrated neuro-symbolic pipeline to generate formally verified data translation code for a MIoT use case. Specifically, we combined the ShinkaEvolve's open-source LLM-based evolutionary algorithm framework with a PVS verification module. Our system uses LLMs to generate translations between JSON schemas using the domain-specific language of PVS. These translations can be formally verified to ensure that the output schema is always followed. We have demonstrated our pipeline via a case study of integrating a pulse oximeter device into our MIoT environment. We are able to consistently generate formally verified translations at low cost.

This works is a first step toward trustworthy automatic integration of novel devices into a existing healthcare infrastructure. In future work we plan to explore integration of a boarder set of devices, formal verification of the networking component of the system, and integrating the formal verification stage into a fully agentic pipeline. 

\section*{Acknowledgements}
This research was, in part, funded by the Advanced Research Projects Agency for Health (ARPA-H). The views and conclusions contained in this document are those of the authors and should not be interpreted as representing the official policies, either expressed or implied, of the United States Government.

\section*{Impact Statement}
This paper presents work whose goal is formally validate LLM generated code for a medical Internet of Things environment. This has the potential impact of making AI-powered medical systems more trustworthy and reliable.

\bibliography{example_paper}
\bibliographystyle{icml2026}

\newpage
\appendix
\onecolumn
\section{Code Listings}
In Listing \ref{lst:input_schema} we display the input schema from the pulse oximeter. In Listing \ref{lst:output_schema} we display the output schema. In Listing \ref{lst:correct_pvs} we display an example of a correctly verified PVS theory that translates the input schema to the output schema. 

\begin{lstlisting}[language=json, caption={Input schema}, label={lst:input_schema}]
{
  "$schema": "https://json-schema.org/draft-07/schema#",
  "additionalProperties": false,
  "properties": {
    "finger_present": {
      "description": "Finger detection indicator",
      "enum": [
        0,
        1
      ],
      "examples": [
        1
      ],
      "type": "integer"
    },
    "pleth": {
      "description": "Plethysmography reading",
      "examples": [
        36
      ],
      "maximum": 99,
      "minimum": 0,
      "type": "integer"
    },
    "pulse_rate": {
      "anyOf": [
        {
          "maximum": 82,
          "minimum": 59
        },
        {
          "const": 255,
          "type": "integer"
        }
      ],
      "description": "Heart rate measurement",
      "examples": [
        79
      ],
      "type": "integer"
    },
    "spo2": {
      "anyOf": [
        {
          "maximum": 100,
          "minimum": 88
        },
        {
          "const": 127,
          "type": "integer"
        }
      ],
      "description": "Blood oxygen saturation percentage",
      "examples": [
        98
      ],
      "type": "integer"
    },
    "valid": {
      "description": "Reading validity indicator",
      "enum": [
        0,
        1
      ],
      "examples": [
        1
      ],
      "type": "integer"
    }
  },
  "required": [
    "pleth",
    "pulse_rate",
    "valid",
    "spo2",
    "finger_present"
  ],
  "type": "object"
}
\end{lstlisting}

\newpage

\begin{lstlisting}[language=json, caption={Output schema}, label={lst:output_schema}]
{
  "$id": "http://hl7.org/fhir/json-schema/6.0",
  "$schema": "http://json-schema.org/draft-07/schema#",
  "description": "See http://hl7.org/fhir/json.html#schema for information about the FHIR JSON Schemas.",
  "properties": {
    "category": {
      "description": "A code that classifies the general type of observation being made.",
      "items": {
        "properties": {
          "coding": {
            "properties": {
              "code": {
                "const": "vital-signs",
                "type": "string"
              },
              "system": {
                "const": "http://terminology.hl7.org/CodeSystem/observation-category",
                "type": "string"
              }
            },
            "required": [
              "system",
              "code"
            ],
            "type": "object"
          }
        },
        "required": [
          "coding"
        ],
        "type": "object"
      },
      "type": "array"
    },
    "code": {
      "properties": {
        "coding": {
          "items": {
            "properties": {
              "code": {
                "const": "2710-2",
                "type": "string"
              },
              "display": {
                "const": "Oxygen saturation in Capillary blood by Oximetry",
                "type": "string"
              },
              "system": {
                "const": "http://loinc.org",
                "type": "string"
              }
            },
            "required": [
              "system",
              "code",
              "display"
            ],
            "type": "object"
          },
          "type": "array"
        },
        "text": {
          "const": "SpO2",
          "type": "string"
        }
      },
      "required": [
        "coding",
        "text"
      ],
      "type": "object"
    },
    "effectiveDateTime": {
      "description": "The time or time-period the observed value is asserted as being true.",
      "type": "string"
    },
    "encounter": {
      "properties": {
        "reference": {
          "description": "Encounter (appointment) ID",
          "type": "string"
        }
      },
      "required": [
        "reference"
      ],
      "type": "object"
    },
    "resourceType": {
      "const": "Observation",
      "type": "string"
    },
    "subject": {
      "properties": {
        "display": {
          "description": "Patient name",
          "type": "string"
        },
        "reference": {
          "description": "Patient ID",
          "type": "string"
        }
      },
      "required": [
        "reference"
      ],
      "type": "object"
    },
    "valueQuantity": {
      "description": "The information determined as a result of making the observation, if the information has a simple value.",
      "properties": {
        "code": {
          "const": "%",
          "type": "string"
        },
        "system": {
          "const": "http://unitsofmeasure.org",
          "type": "string"
        },
        "unit": {
          "const": "%",
          "type": "string"
        },
        "value": {
          "description": "The information determined as a result of making the observation, if the information has a simple value.",
          "type": "number"
        }
      },
      "required": [
        "value"
      ],
      "type": "object"
    }
  },
  "required": [
    "resourceType",
    "category",
    "code",
    "subject",
    "effectiveDateTime",
    "valueQuantity"
  ],
  "type": "object"
}
\end{lstlisting}

\newpage

\begin{lstlisting}[language=pvs, caption={Verified Correct PVS Theory}, label={lst:correct_pvs}]
A_B: THEORY
BEGIN

  jsltlib: LIBRARY = "../../jslt_pvs/jsltlib/"
    
  IMPORTING jsltlib@jsondata

  transform(i: jsondata): jsondata =
    LET spo2v: jsondata = json_get_key(i, "spo2"),
        mk = (LAMBDA (k: string, v: jsondata): jpair(k, v))
    IN
      jdict((#
        length := 6,
        seq := [:
          mk("resourceType", jstr("Observation")),
          mk("category",
            jarray((#
              length := 1,
              seq := [:
                jdict((#
                  length := 1,
                  seq := [:
                    mk("coding",
                      jdict((#
                        length := 2,
                        seq := [:
                          mk("system", jstr("http://terminology.hl7.org/CodeSystem/observation-category")),
                          mk("code", jstr("vital-signs"))
                        :]
                      #))
                    )
                  :]
                #))
              :]
            #))
          ),
          mk("code",
            jdict((#
              length := 2,
              seq := [:
                mk("coding",
                  jarray((#
                    length := 1,
                    seq := [:
                      jdict((#
                        length := 3,
                        seq := [:
                          mk("system", jstr("http://loinc.org")),
                          mk("code", jstr("2710-2")),
                          mk("display", jstr("Oxygen saturation in Capillary blood by Oximetry"))
                        :]
                      #))
                    :]
                  #))
                ),
                mk("text", jstr("SpO2"))
              :]
            #))
          ),
          mk("subject",
            jdict((#
              length := 1,
              seq := [:
                mk("reference", jstr("Patient/example"))
              :]
            #))
          ),
          mk("effectiveDateTime", jstr("1970-01-01T00:00:00Z")),
          mk("valueQuantity",
            jdict((#
              length := 4,
              seq := [:
                mk("value", spo2v),
                mk("unit", jstr("%")),
                mk("system", jstr("http://unitsofmeasure.org")),
                mk("code", jstr("%"))
              :]
            #))
          )
        :]
      #))

END A_B
\end{lstlisting}

\end{document}